\documentclass{emulateapj}
\newcommand{\degree}{\hbox{$^\circ$}}
\newcommand{\etal}{et\,al.}
\newcommand{\halpha}{H$\alpha$}
\newcommand{\gsim}{\raise0.3ex\hbox{$>$}\kern-0.75em{\lower0.65ex\hbox{$\sim$}}}
\newcommand{\kms}{km\,s$^{-1}$}
\newcommand{\lsim}{\raise0.3ex\hbox{$<$}\kern-0.75em{\lower0.65ex\hbox{$\sim$}}}

\newcommand{\msun}{M$_{\odot}$}      

\newcommand{\pom}{\,$\pm$\,}

\newcommand{\HI}{H~{\sc i}}

\newcommand{\adbs}{ADBS\,1138}
\newcommand{\vla}{{\it VLA}}
\newcommand{\mjypbm}{mJy\,Bm$^{-1}$}
\newcommand{\jykms}{Jy\,km\,s$^{-1}$}
\begin{document}     
\slugcomment{{\rm Astrophysical Journal, in press}}

%-----------------------------------------------------------------------------%
\title{Quiescent Isolation: The Extremely Extended HI Halo of the Optically
Compact Dwarf Galaxy ADBS 113845+2008}
%-----------------------------------------------------------------------------%

%------------------------------
%Author list - emulateapj style
%------------------------------
\author{John M. Cannon\altaffilmark{1,2},
John J. Salzer\altaffilmark{3,2},
Jessica L. Rosenberg\altaffilmark{4}}

\altaffiltext{1}{Department of Physics \& Astronomy, Macalester College, 
1600 Grand Avenue, Saint Paul, MN 55105; jcannon@macalester.edu}
\altaffiltext{2}{Astronomy Department, Wesleyan University, 
Middletown, CT 06457}
\altaffiltext{3}{Department of Astronomy, Indiana University, 
727 East Third Street, Bloomington, IN 47405; slaz@astro.indiana.edu}
\altaffiltext{4}{Department of Physics \& Astronomy, George Mason University,
Fairfax, VA 22030; jrosenb4@gmu.edu}

%---------------------------
%Author list - journal style 
%---------------------------
				   
%\author{John M. Cannon}
%\affil{Department of Physics \& Astronomy, Macalester College, 1600 Grand Avenue, Saint Paul, MN 55105}
%\affil{Astronomy Department, Wesleyan University, Middletown, CT 06457}
%\email{jcannon@macalester.edu}

%\author{John J. Salzer}
%\affil{Department of Astronomy, Indiana University, 727 East Third Street, 
%Bloomington, IN 47405}
%\affil{Astronomy Department, Wesleyan University, Middletown, CT 06457}
%\email{slaz@astro.indiana.edu}

%\author{Jessica L. Rosenberg}
%\affil{Department of Physics \& Astronomy, George Mason University,
%Fairfax, VA 22030}
%\email{jrosenb4@gmu.edu}

%-----------------------------------------------------------------------------%
\begin{abstract}
%-----------------------------------------------------------------------------%

We present new optical imaging and spectroscopy and \HI\ spectral line
imaging of the dwarf galaxy ADBS\,113845$+$2008 (hereafter \adbs).
This metal-poor (Z $\sim$30\% Z$_{\odot}$), ``post-starburst'' system
has one of the most compact stellar distributions known in any galaxy
to date (B-band exponential scale length $=$0.57 kpc).  In stark
contrast to the compact stellar component, the neutral gas is
extremely extended; \HI\ is detected to a radial distance of $\sim$25
kpc at the 10$^{19}$ cm$^{-2}$ level ($\gsim$44 B-band scale lengths).
Comparing to measurements of similar ``giant disk'' dwarf galaxies in
the literature, \adbs\ has the largest known \HI-to-optical size
ratio.  The stellar component is located near the center of a broken
ring of \HI\ that is $\sim$15 kpc in diameter; column densities peak
in this structure at the $\sim$ 3.5\,$\times$\,10$^{20}$ cm$^{-2}$
level.  At the center of this ring, in a region of comparatively low
\HI\ column density, we find ongoing star formation traced by \halpha\
emission.  We sample the rotation curve to the point of turn over;
this constrains the size of the dark matter halo of the galaxy, which
outweighs the luminous component (stars $+$ gas) by at least a factor
of 15.  To explain these enigmatic properties, we examine
``inside-out'' and ``outside-in'' evolutionary scenarios.
Calculations of star formation energetics indicate that ``feedback''
from concentrated star formation is not capable of producing the ring
structure; we posit that this is a system where the large \HI\ disk is
evolving in quiescent isolation.  In a global sense, this system is
exceedingly inefficient at converting neutral gas into stars.

\end{abstract}						


%-----------------------------------------------------------------------------%
\section{Introduction}
\label{S1}
%-----------------------------------------------------------------------------%

Blue compact dwarf (BCD) galaxies are intrinsically faint, (e.g.,
M$_{\rm B} \gsim -$17) compact (optical radii of a few kpc or smaller)
systems that are undergoing strongly concentrated active star
formation.  Their interstellar media typically produce strong nebular
emission lines that suggest sub-solar abundances, as is common for
low-mass galaxies in a general sense
\citep[e.g.,][]{skillman89,salzer05,lee06}.  The active star formation
regions are typically superposed on an underlying, lower-surface
brightness red stellar population.  Thus, as expected, BCDs have
higher central surface brightnesses than do more quiescent dwarf
irregular (dIrr) galaxies.  Further, the underlying stellar
populations in BCDs are also more compact than those in dIrr's
\citep{salzer99}, the optical disk scale lengths being systematically
shorter.

As expected from their elevated star formation rates, most BCDs
contain plentiful neutral gas.  When compared with more quiescent dIrr
galaxies of similar luminosity, BCDs typically have \HI\ masses that
are roughly twice as large \citep{salzer02}.  Like the stellar
component, this \HI\ gas is in general more centrally concentrated
than in the dIrr population.  The \HI\ distributions are often
strongly peaked at or near the location of the starburst region
\citep{vanzee98}.

These properties suggest that BCDs are simply small galaxies with
centrally-concentrated mass distributions.  Thus, the growing number
of low-mass galaxies with extended \HI\ disks
\citep[e.g.,][]{carignan88,bajaja94,young96,hunter98a,vanzee04,warren04,begum05}
might be simplistically interpreted as a collection of relatively
quiescent dIrr systems where the mass is more evenly distributed
throughout the disk than in BCDs.  Likewise, a centrally-concentrated
\HI\ or stellar distribution might be the signpost of a BCD.

Strong departures from this simple paradigm would be found in star
forming galaxies that have extreme stellar and gaseous properties: the
compact stellar bodies typical of BCDs and the extended \HI\ gas disks
typical of more quiescent irregular galaxies.  While examples of BCDs
with large \HI\ envelopes exist (e.g., NGC\,2915; {Meurer \etal\
1994}\nocite{meurer94}, {1996}\nocite{meurer96}), we do not yet have
sufficient statistics on the resolved characteristics of the gaseous
components of BCD systems to determine if such remarkable properties
are common or rare.  Further, few systems with extremely compact
stellar populations (optical scale lengths $\lsim$1 kpc) have been
identified to date; such stellar distributions are exceptional even
among the BCD class.

\begin{figure*}
\plotone{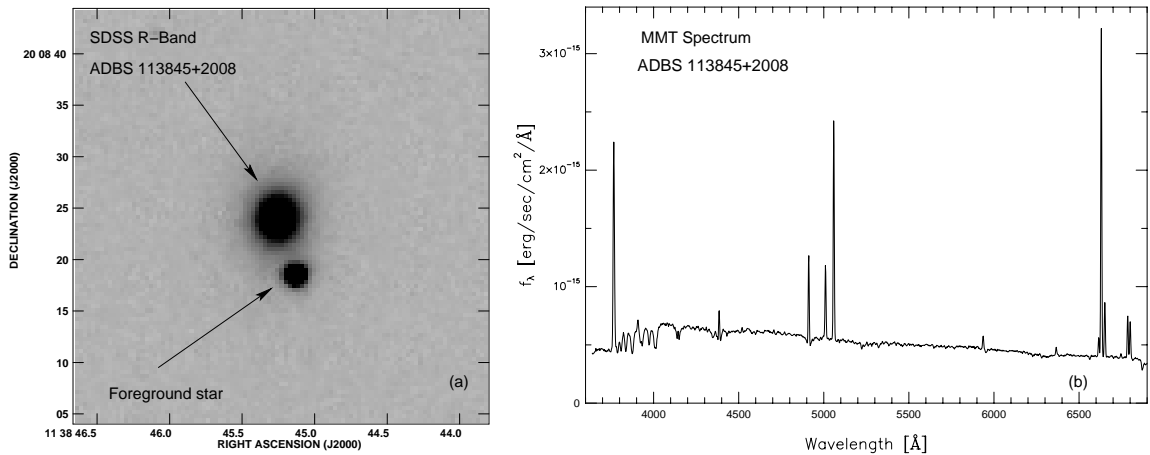}
\caption{{\it Sloan Digitized Sky Survey} ({\it SDSS}) r-band image
{\it (a)}, showing separation of \adbs\ and a nearby foreground star;
note the extremely compact optical component of the galaxy.  {\it MMT}
optical spectrum {\it (b)}, demonstrating ``post-starburst''
properties of the target galaxy.}
\label{figcap1}
\end{figure*}

To explore this uncharted territory, we present in this work new
optical and \HI\ spectral line observations of ADBS113845$+$2008
(hereafter \adbs), an extremely compact BCD surrounded by an enormous
\HI\ disk.  Initially discovered in the {\it Arecibo Dual Beam Survey}
(ADBS) \citep{rosenberg00,rosenberg02}, it was subsequently cataloged
as a UV-excess source in the 2nd KISO survey ({Miyauchi-Isobe \&
Maehara, 2000}\nocite{miyauchi00}).  Its extremely compact nature was
first recognized during a CCD imaging study of the ADBS galaxies (see
section 2.1), which motivated us to carry out the more detailed study
of this interesting system presented here.  In addition to describing
our new optical and radio data, this paper discusses the morphology
and kinematics of the neutral gas, the nature of the \HI\ ring and
central depression, and different scenarios that may produce the
curious features of the ISM in \adbs.

%-----------------------------------------------------------------------------%
\section{Observations and Data Reduction}
\label{S2}
%-----------------------------------------------------------------------------%

%-----------------------------------------------------------------------------%
\subsection{Optical Imaging and Spectroscopy}
\label{S2.1}
%-----------------------------------------------------------------------------%

\begin{figure*}
\plotone{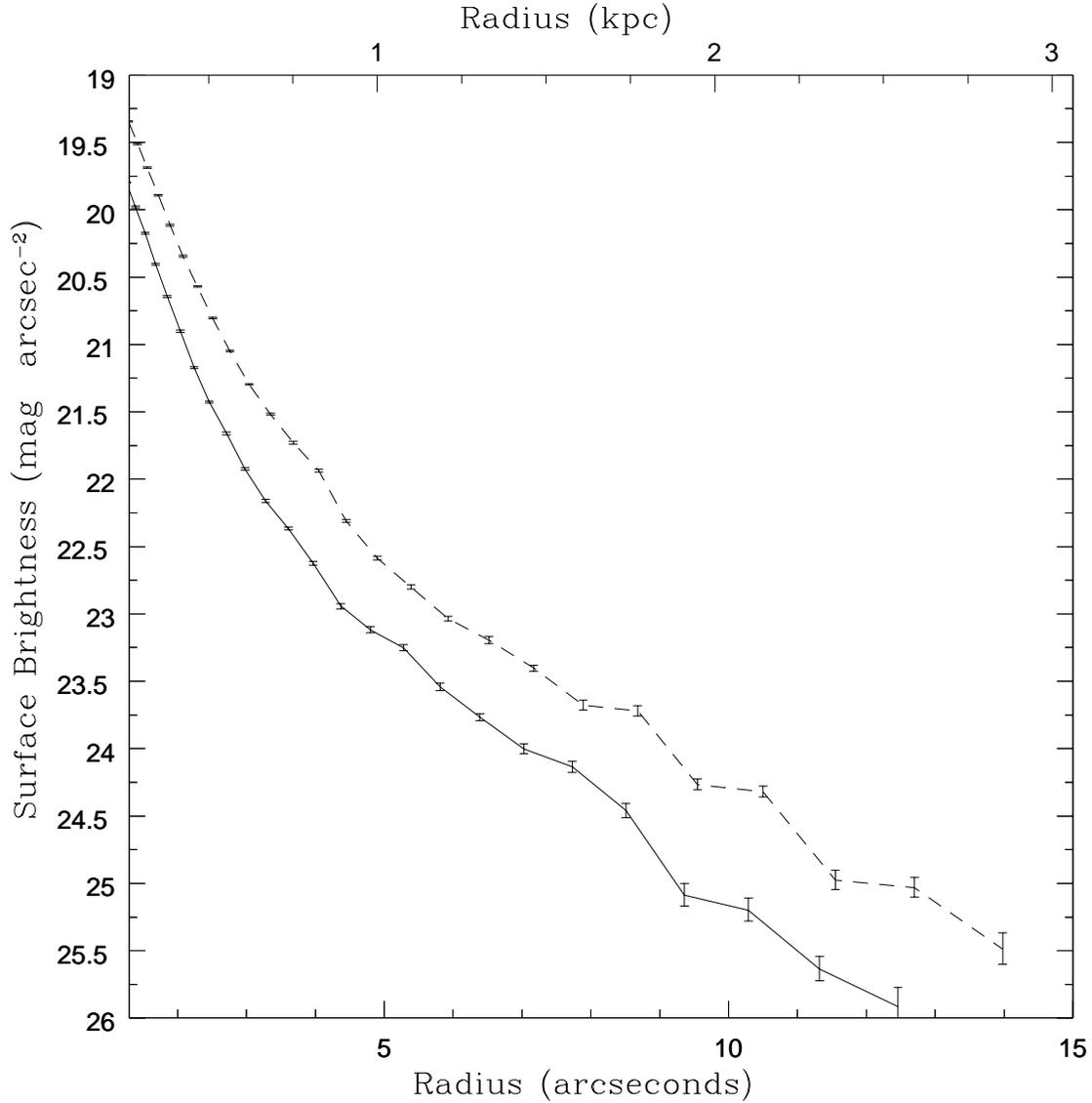}
\caption{B-band (solid line) and R-band (dotted line) radial profiles
of \adbs\ from {\it WIYN} 0.9-meter images.  These profiles were
created by fitting elliptical isophotes to the galaxy.  The fits give
exceptionally short exponential scale lengths (e.g., 2.79\arcsec\ $=$
0.57 kpc in the B-band).}
\label{figcap2}
\end{figure*}

Optical imaging of \adbs\ through B, V and R filters was carried out
using the {\it WIYN} 0.9-meter telescope\footnote{The {\it WIYN}
0.9-meter telescope is operated jointly by a consortium that includes
the University of Florida, Indiana University, San Francisco State
University, Wesleyan University, University of Wisconsin-Madison,
University of Wisconsin-Oshkosh, University of Wisconsin-Stevens
Point, and University of Wisconsin-Whitewater} located on Kitt Peak.
The data were obtained on 5 May 2003 as part of a larger program to
study the optical properties of the full ADBS sample (Stevenson \etal,
in preparation).  The exceptionally compact nature of this system
was evident immediately.  In fact, in our initial analysis of the CCD
images we didn't recognize \adbs\ as a galaxy at all, but rather took
it to be a foreground star.  Only after manipulating the image
intensity and measuring the FWHM of several objects did we identify it
as a galaxy.  Figure~\ref{figcap1}(a), obtained from the Sloan Digital
Sky Survey \citep{adelman08}, shows the immediate field surrounding
the galaxy; note that it is only slightly more extended than the
nearby foreground star to the southwest.

The measured B-band apparent magnitude of \adbs\ is 16.51.  Assuming a
distance of 42.1 Mpc (H$_0$ = 73 km\,s$^{-1}$\,Mpc$^{-1}$), the
absolute magnitude of \adbs\ (corrected for foreground extinction of
0.097 mag) is M$_{\rm B}$ $-$16.71.  This corresponds to L$_{\rm B}$ =
7.44\,$\times$\,10$^8$ L$_{\odot}$ (assuming M$_{{\rm B},\odot} =
+$5.47; see {Cox 2000}\nocite{cox00}).  The system has a B-band
exponential scale length of only $\sim$0.57 kpc (see
Figure~\ref{figcap2}).  This can be compared to a typical scale length
of 1.54\,$\pm$\,0.22 kpc for other dwarfs in the ADBS sample with
similar luminosities.  Examination of both our {\it WIYN} 0.9-meter
images and the {\it SDSS} data reveal a compact, featureless
morphology.  Furthermore, the galaxy appears to be quite isolated; no
optical or \HI-rich objects with a comparable redshift are located
within $\sim$16\arcmin\ of \adbs\ (half of the \HI\ primary beam at
21\,cm; see Section~\ref{S2.2}), corresponding to $\sim$195 kpc at the
adopted distance.

We also obtained \halpha\ narrow-band imaging with the {\it WIYN}
0.9-m telescope on 2 April 2005.  The \halpha\ emission from \adbs\
comes solely from the compact core of the galaxy, and is unresolved in
our images.  The measured flux derived from our image is
2.98\,$\times$\,10$^{-14}$ erg\,s$^{-1}$\,cm$^{-2}$.  After applying
modest corrections for foreground Galactic extinction (A$_{\rm R} =$
0.06 mag, {Schlegel \etal\ 1998}\nocite{schlegel98}), internal
absorption in the galaxy, and [N II] nebular emission present in our
narrow-band filter, we arrive at a corrected \halpha\ flux of
3.65\,$\times$\,10$^{-14}$ erg\,s$^{-1}$\,cm$^{-2}$.  At our adopted
distance this corrected flux corresponds to a luminosity of
7.75\,$\times$\,10$^{39}$ erg\,s$^{-1}$.  Applying the conversion to
SFR from \citet{kennicutt98a}, \adbs\ has a current SFR of 0.06
\msun\,yr$^{-1}$.

Optical spectroscopy of \adbs\ was obtained with the {\it MMT} 6.5\,m
telescope on 5 February 2006\footnote{The MMT Observatory is a joint
facility of the Smithsonian Institution and the University of Arizona}
using the Blue Channel Spectrograph and a 300 l/mm grating.  A slit
width of 1.5\arcsec\ was employed, and the total integration time was
30 minutes, split between two exposures.  Data processing followed
standard practices; the wavelength scale was assigned using exposures
of a HeNeAr lamp, and the flux scale was obtained by measurement of
several spectrophotometric standard stars.  The resulting spectrum,
shown in Figure~\ref{figcap1}(b), allows us to estimate the nebular
abundance of ADBS\,1138 to be $\sim$30\% Z$_{\odot}$, using the
``strong-line'' method of \citet{salzer05}.  We measure an optical
recession velocity of 3076\,$\pm$\,14 \kms.  \halpha\ and [\ion{O}{3}]
emission lines are prominent; the spectrum also reveals Ca~II~K, Mg~b,
and Na~D absorption features, and considerable absorption in the
higher-order Balmer series transitions.  Taken together, these
features indicate the presence of a substantial population of older
stars \citep[e.g.,][]{worthey97} and imply a higher SFR in the past
than at the present epoch (i.e., that the system is in a
``post-starburst'' state).  This important clue is discussed further
in the sections that follow.  We summarize salient properties of
\adbs\ in Table~\ref{t1}.

\begin{deluxetable*}{ccc}
\tabletypesize{\scriptsize}
\tablecaption{Basic Parameters of ADBS 113845$+$2008}
\tablewidth{0pt}
\tablehead{
\colhead{Property}         
&\colhead{Value} 
&\colhead{Reference}}
\startdata
V$_{{\rm HI}}$ (\kms)	        &3074\,$\pm$\,1 		&This work\\
V$_{\rm opt}$ (\kms)	        &3076\,$\pm$\,14 		&This work\\
Distance (Mpc)\tablenotemark{a} &42.1\,$\pm$\,4	        	&This work\\
M$_{\rm B}$			&$-$16.71		        &This work\\
(B$-$V)                         &0.59\,$\pm$\,0.13              &This work\\
Gal. Lat. (\degree)	        &72.0    		        &$--$\\
Foreground E(B$-$V)             &0.022	                        &{Schlegel \etal\ (1998)}\nocite{schlegel98}\\
12\,$+$\,log(O/H) 	        &8.3\,$\pm$\,0.3	       	&This work\\
Current SFR (\msun\,yr$^{-1}$)  &0.06  		                &This work\\
\HI\ Mass (10$^9$ \msun)        &1.08\,$\pm$\,0.1		&This work\\
Dynamical Mass\tablenotemark{b} (10$^{10}$ \msun)  &3.8\,$\pm$\,0.8         &This work\\
\enddata
\tablenotetext{a}{Assumes H$_0$ = 73 km\,s$^{-1}$\,Mpc$^{-1}$.}
\tablenotetext{b}{Measured at the last reliable point of the rotation curve.}
\label{t1}
\end{deluxetable*}

%-----------------------------------------------------------------------------%
\subsection{\HI\ Spectral Line Imaging}
\label{S2.2}
%-----------------------------------------------------------------------------%

The {\it Very Large Array} (\vla\footnote{The National Radio Astronomy
Observatory is a facility of the National Science Foundation operated
under cooperative agreement by Associated Universities, Inc.})  was
used to obtain \HI\ spectral line data for \adbs\ as part of observing
program AC\,841.  Observations were obtained in the C and D arrays on
2006 October 31 and 2007 March 30, respectively; integration times
were $\sim$243 and $\sim$87 minutes for the target source during these
two observing sessions.  A total bandwidth of 1.56 MHz was used, with
128 channels separated by 12.2 kHz (2.58 \kms).  The data were reduced
and calibrated using standard methods in the AIPS environment;
analysis was performed using the AIPS, Miriad and GIPSY packages.

Two \HI\ datacubes are created and analyzed after off-line Hanning
smoothing to a velocity resolution of 5.2 \kms.  The
``low-resolution'' cube was created with natural weighting (ROBUST=5
in the AIPS IMAGR task) and then convolved to a circular beam size of
40\arcsec.  The ``high-resolution'' cube was created with robust
weighting (ROBUST=0.5) and then convolved to a circular beam size of
20\arcsec.  The rms noise in the final cubes are 0.35 and 0.29
\mjypbm\ for the natural and robust-weighted cubes, respectively.
Following \citet{jorsater95} and \citet{walter99}, we explicitly
correct for the difference in areas between the the clean and dirty
beams, resulting in accurate \HI\ flux calibration of our
interferometric data.

We create moment maps of the integrated \HI\ emission and velocity
structure using an approach similar to that advocated by
\citet{walter97}.  First, the ``low-resolution'' cube was blanked at
the 2\,$\sigma$ level.  Channels containing line emission were then
inspected individually, and we differentiate real from spurious
emission by requiring structures to be present in 3 or more
consecutive channels.  This edited cube is then used to blank both the
``low-'' and ``high-resolution'' cubes, assuring that the same regions
contribute to both final moment maps.  The images of integrated \HI\
emission, discussed in detail in \S~\ref{S3}, probe column densities
of $\gsim$10$^{19}$ cm$^{-2}$.

During the reduction of the \HI\ spectral line data, we averaged 19
line-free channels to produce a 1.4 GHz radio continuum image.  This
image has a beam size of 31\arcsec\,$\times$\,25\arcsec\ and an RMS
noise level of 0.27 \mjypbm.  The 5$\sigma$ upper limit to the 1.4 GHz
radio continuum flux density is derived to be S$_{\rm 1.4 GHz}
\lsim$1.3 mJy.  This is consistent with the NVSS non-detection
\citep{condon98}.  The expected thermal radio continuum flux density,
based on the (foreground extinction-corrected) \halpha\ flux, is
between 0.04--0.06 mJy (depending on the adopted electron temperature;
see {Caplan \& Deharveng 1986}\nocite{caplan86}).  Thus, even if
nonthermal processes dominate the radio continuum flux, it is below
our current sensitivity level to study in detail.

\begin{figure*}
\plotone{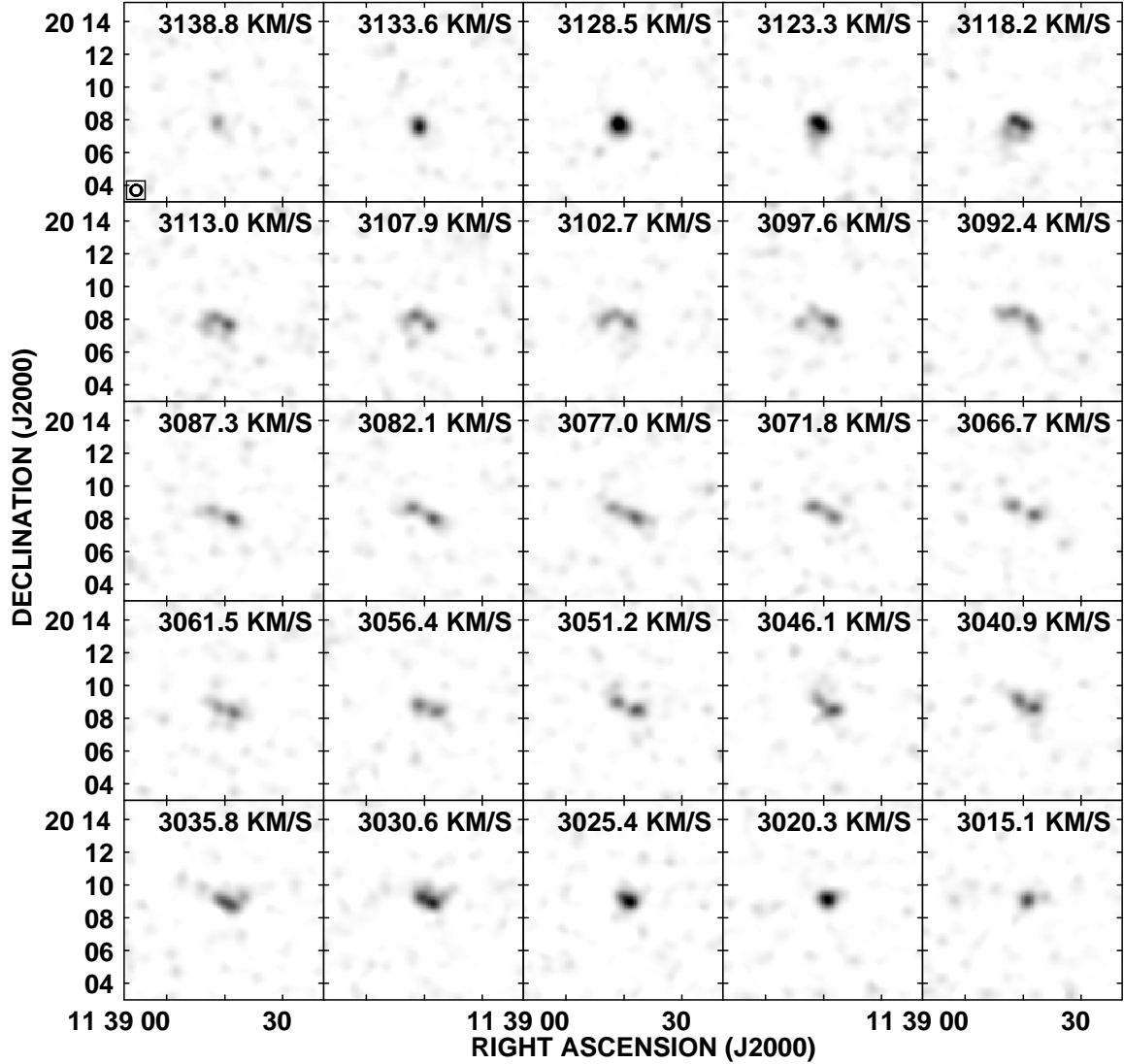}
\caption{Individual channel maps of \HI\ emission in the
``low-resolution'' (40\arcsec\ beam) datacube.  The ring structure and
central \HI\ depression are both prominent (consider velocities from
$\sim$3092 -- 3050 \kms).}
\label{figcap3}
\end{figure*}

%-----------------------------------------------------------------------------%
\section{\HI\ Emission in ADBS 113845+2008}
\label{S3}
%-----------------------------------------------------------------------------%

%-----------------------------------------------------------------------------%
\subsection{\HI\ Distribution}
\label{S3.1}
%-----------------------------------------------------------------------------% 

\HI\ in \adbs\ is detected over the (Heliocentric) velocity range
3010$-$3140 \kms; Figure~\ref{figcap3} shows channel maps of \HI\
emission in the central 25 channels of the cube.  From these maps it
is clear that the profile is double-peaked; this is verified when all
emission defined as real (i.e., all emission contributing to the
moment maps) is summed and plotted against velocity in the global \HI\
profile shown in Figure~\ref{figcap4}.  From this profile and rotation
curve analysis (see below), we derive a systemic velocity of V$_{\rm sys}
=$ 3074\,$\pm$\,1 \kms\ which is in excellent agreement with the
measured optical velocity (see \S~\ref{S2.1}).

Integrating under the \HI\ profile, we derive a total \HI\ flux S $=$
2.58\,$\pm$0.26 \jykms.  This can be compared to the lower value of 1.98
\jykms\ derived from {\it Arecibo} observations for the ADBS
\citep{rosenberg00}.  This discrepancy is easily explained as the \HI\
source passing off the center of the {\it Arecibo} beam during the
drift scan observations.  The slightly lower systemic and optical
velocities derived here (3074 and 3076 \kms, respectively) compared to
the value from the ADBS (3105 \kms) agrees with this interpretation.
Note that such disagreement between single-dish and interferometric
observations were common in the ADBS sample; of the 31 galaxies 
observed by \citet{rosenberg00} with both the {\it VLA} and {\it Arecibo},
85\% have higher interferometric than single-dish \HI\ fluxes.  At the 
adopted distance of 42.1 Mpc, this flux integral translates to a total 
\HI\ mass of (1.08\,$\pm$\,0.22)\,$\times$\,10$^9$ \msun.  Combining with 
the B-band luminosity (7.44\,$\times$\,10$^8$ L$_{\odot}$) implies 
M$_{\rm H}$/L$_{\rm B}$ $\sim$1.45.

\begin{figure*}
\plotone{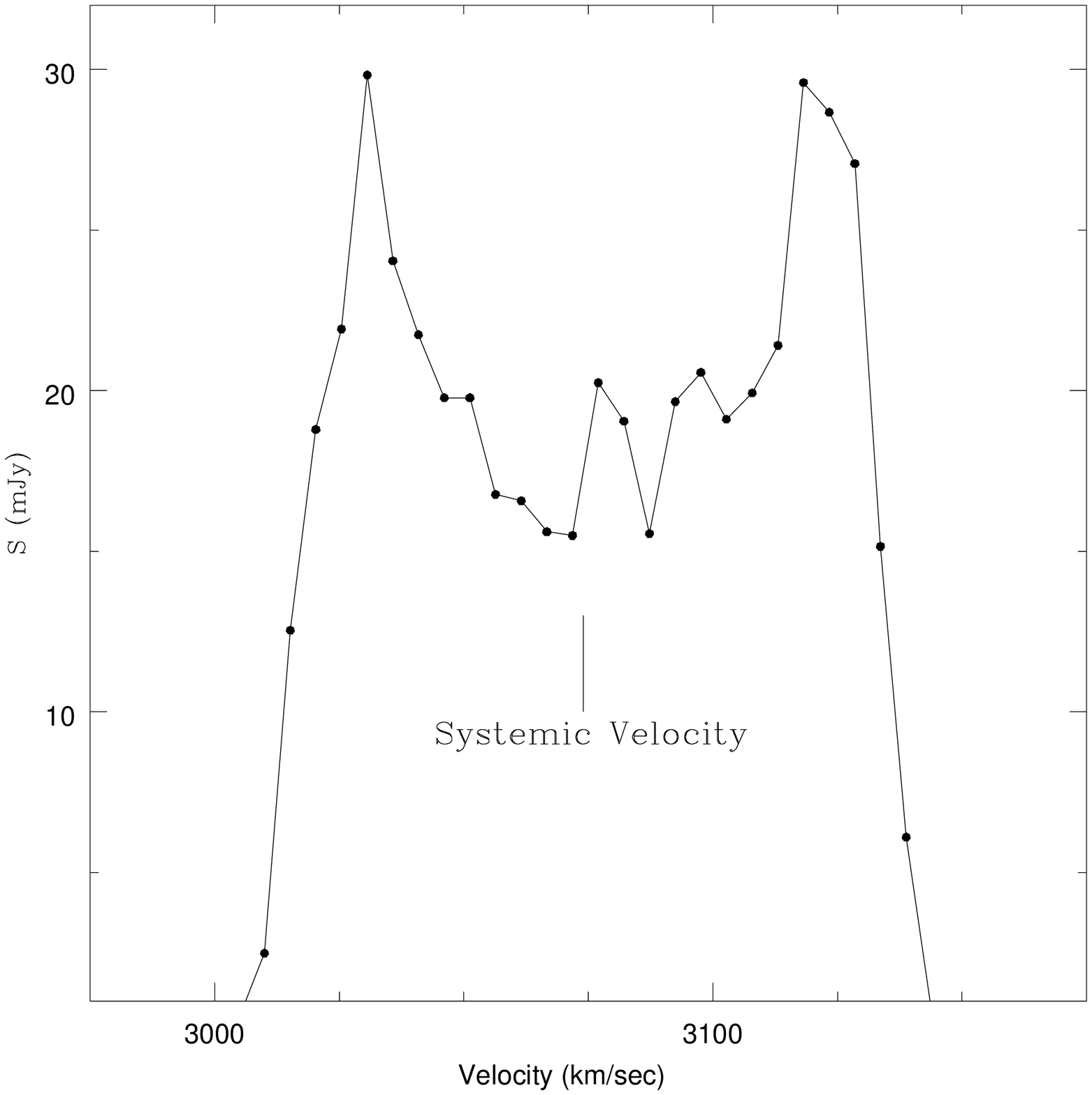}
\caption{The global \HI\ profile of ADBS\,1138, calculated by summing
the flux densities in individual channels of the ``low-resolution''
(40\arcsec\ beam) datacube.  The adopted systemic velocity, 3074\pom1
km\,s$^{-1}$, is in excellent agreement with the optical velocity (see
Figure~\ref{figcap1}).}
\label{figcap4}
\end{figure*}

\begin{figure*}
\plotone{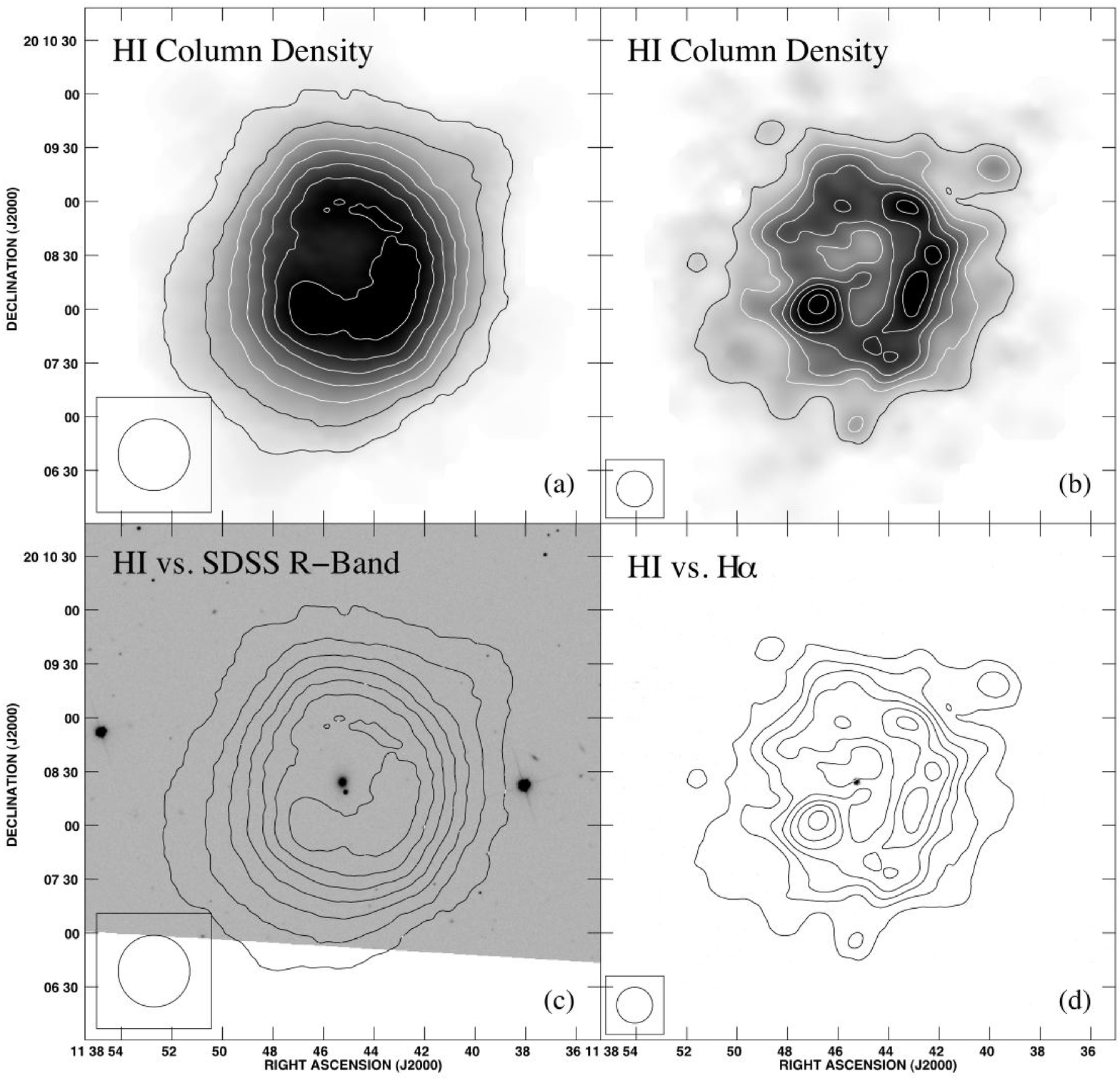}
\caption{Contours of \HI\ column density overlaid on images of the
\HI\ column density distribution (40\arcsec\ resolution - panel {\it
a}; 20\arcsec\ resolution - panel {\it b}), the {\it SDSS} r-band
image (panel {\it c}), and the {\it WIYN} 0.9-m \halpha\ image (panel
{\it d}).  Contours in the 40\arcsec\ resolution panels are at levels
of (0.5, 1.0, 1.5, 2.0, 2.5, 3.0, 3.5)\,$\times$\,10$^{20}$ cm$^{-2}$;
contours in the 20\arcsec\ resolution panels are at levels of (0.2,
0.5, 0.8, 1.1, 1.4, 1.7, 2.0)\,$\times$\,10$^{20}$ cm$^{-2}$.  Note
the striking disparity between the extent of \HI\ and the extremely
compact optical component; the ongoing star formation, concentrated in
the very compact optical component, is not associated with high column
density gas at these physical resolutions.}
\label{figcap5}
\end{figure*}

In Figure~\ref{figcap5} we compare {\it SDSS} r-band and {\it WIYN}
0.9-m \halpha\ emission-line images of \adbs\ to \HI\ moment 0 images
at ``low-'' (40\arcsec\ beam) and ``high-resolution'' (20\arcsec\
beam).  It is immediately striking that neutral gas extends
$\gsim$125\arcsec\ ($\gsim$25 kpc) from the optical component.
Comparing to the B-band scale length ($\sim$0.57 kpc), the \HI\ disk is
$\sim$44 times larger than the B-band exponential scale length.  

Comparing to other ``giant disk'' dwarf galaxies in the literature, we
conclude that \adbs\ is among the most extended \HI\ systems known.
Due to the very compact nature of \adbs, and to potential
contamination of a surface brightness estimate (e.g., when measuring
the Holmberg radius) due to the bright foreground star in the
southwest (see Figure~\ref{figcap1}), we choose for this comparison
the ratio of \HI\ size at the 10$^{19}$ cm$^{-2}$ level to the B-band
exponential scale length.  We denote this parameter as
$\mathcal{R}$(\HI/B); as shown in Table~\ref{t2}, other ``giant disk''
dwarf galaxies have similar, but slightly smaller values of this
parameter.  Each of these systems shows extreme properties when 
comparing the stellar and gaseous components. 

The optical scale length of \adbs\ does not vary appreciably with
wavelength in the {\it SDSS} bands (see also Figure~\ref{figcap2}.
This suggests that we are likely observing the entire stellar
component at these wavelengths.  If the galaxy harbors an extended,
low-surface brightness disk, then this component must have a surface
brightness well below the sensitivity of our optical images (26.0,
25.5 mag arcsec$^{-2}$ in the B and V bands, respectively).  As a
point of reference we note that deep optical imaging of 27 compact
emission-line galaxies (which are a class of objects that are brighter
than ADBS 1138) by \citet{barton06} finds extended, low surface
brightness emission in only 1 case.

The highest column density gas in \adbs\ is located in a broken ring
structure that is significantly beyond the optical component.  Moving
outward from the dynamical center (which corresponds closely with the
stellar component; see detailed discussion in \S~\ref{S3.2}),
Figure~\ref{figcap6} shows that the \HI\ surface brightness peaks at a
radius of $\sim$40\arcsec\ (8 kpc).  The \HI\ surface brightness falls
by nearly 50\% moving inward toward the optical component.  Therefore,
the shell is $\gsim$15 kpc in diameter, making it one of the largest
such features known in any dwarf galaxy.  The mass of gas in the \HI\
ring, integrated radially from 5 to 12 kpc, is
(5.2$\pm$0.6)\,$\times$\,10$^{8}$ \msun.  Thus, this structure
accounts for roughly half of the total \HI\ mass in \adbs.

Equally remarkable is the modest strength of the integrated \HI\
emission.  Examining the ``high-resolution'' column density map in
Figure~\ref{figcap5}, \HI\ column densities peak at values of
3.8\,$\times$\,10$^{20}$ cm$^{-2}$ and fall below
2\,$\times$\,10$^{20}$ cm$^{-2}$ at the location of the optical
component of the galaxy.  A number of authors have demonstrated a
correlation between \HI\ column density and ongoing star formation;
the canonical ``Schmidt star formation law'' suggests that \HI\
columns of $\sim$10$^{21}$ cm$^{-2}$ are strongly correlated with
instantaneous star formation tracers, including \halpha\ (see
{Skillman 1987}\nocite{skillman87}, {Kennicutt
1989}\nocite{kennicutt89}, {1998b}\nocite{kennicutt98b}, and
references therein).

The \halpha\ image in Figure~\ref{figcap5} shows that \adbs\ has
spatially concentrated ongoing star formation (0.06 \msun\,yr$^{-1}$).
\halpha\ emission is associated with recent massive star formation
(i.e., stars with masses $\gsim$10 \msun\ and main sequence lifetimes
$\lsim$20 Myr; see {Kennicutt 1998a}\nocite{kennicutt98a}).  However,
at the physical resolutions of our \HI\ data, we find star formation
{\it only} in the central, low-column density region.  We discuss the
constraints our data offer on this puzzle in \S~\ref{S4}.

%-----------------------------------------------------------------------------%
\subsection{\HI\ Kinematics}
\label{S3.2}
%-----------------------------------------------------------------------------% 

\begin{figure*}
\plotone{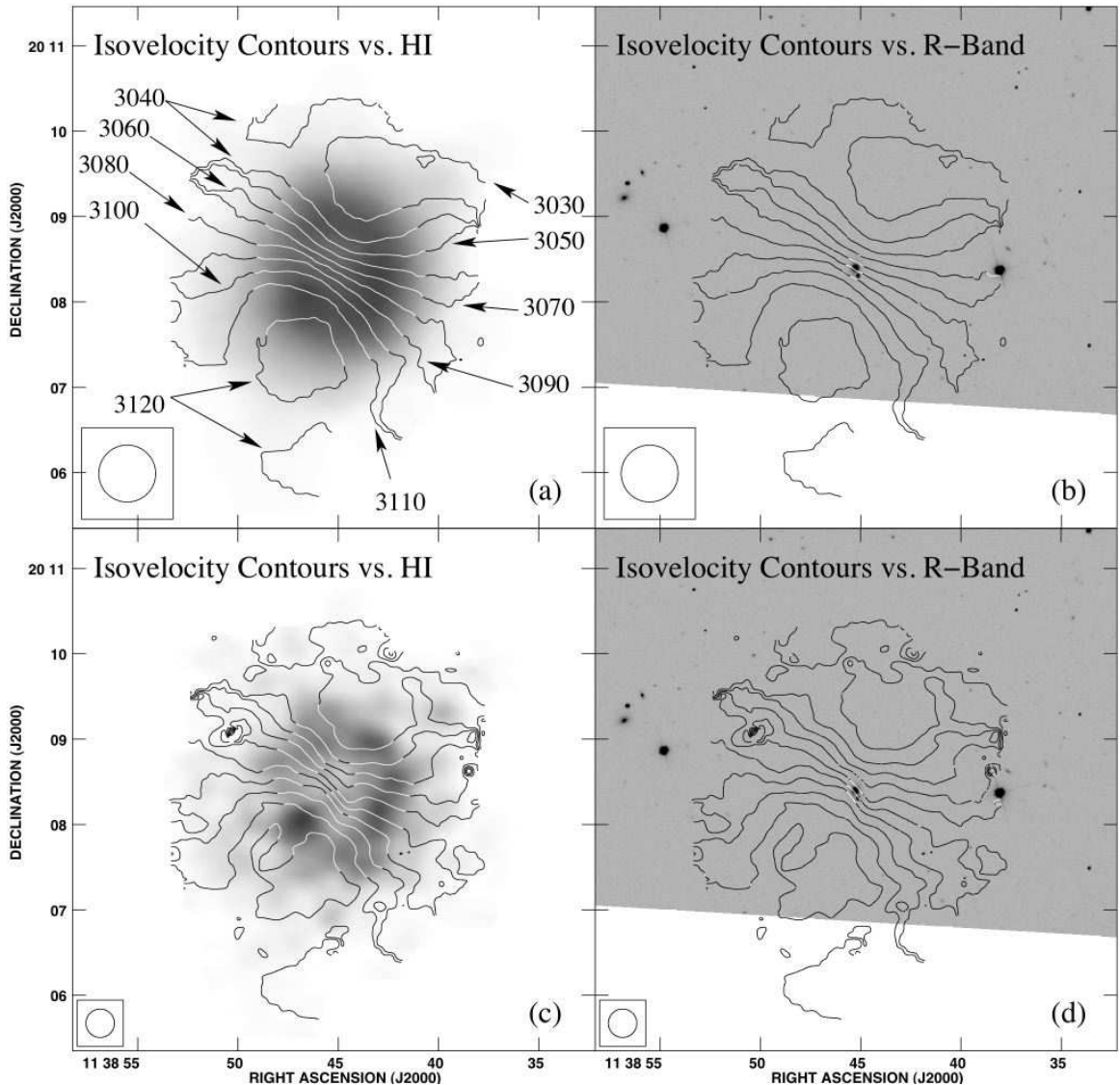}
\caption{Neutral gas isovelocity contours at 40\arcsec\ resolution
(panels {\it a}, {\it b}) and at 20\arcsec\ resolution (panels {\it
c}, {\it d}), overlaid on grayscale representations of the \HI\ column
density images (40\arcsec\ resolution in {\it a}, 20\arcsec\
resolution in {\it c}) and on the {\it SDSS} r-band image (panels {\it
b}, {\it d}).  While the \HI\ morphology is dominated by the prominent
ring, low column density \HI\ gas is present in the central regions;
the system is undergoing ordered rotation within the inner disk.  From
these velocity fields we extract the rotation curves discussed in
\S~\ref{S3.2} and shown in Figure~\ref{figcap8}.}
\label{figcap7}
\end{figure*}

In Figure~\ref{figcap7} we present the velocity structure of the
\adbs\ system at 20\arcsec\ and 40\arcsec\ resolution.  A general
inspection shows that the system is undergoing well-ordered rotation
throughout most of the inner disk (but note that the slope of the
curve changes at our physical resolutions in the inner regions of the
galaxy).  Isovelocity contours appear evenly spaced and are mostly
regular along the \HI\ minor axis.

We fit this velocity field using tilted ring models via the ROTCUR
task in GIPSY.  While the size of the galaxy compared to our smaller
beam size (20\arcsec) does not lend itself to detailed rotation curve
analysis with large numbers of independent points, we can extract a
clean first-order measurement of the large-scale kinematics of the
neutral gas.  ROTCUR fits the systemic velocity (V$_{\rm sys}$), rotation
velocity, position angle (P.A., measured as positive when rotating
eastward from north), inclination {\it i}, and dynamical center
position, each as a function of radius, to an observed velocity field.

We began by fitting the inclination of the galaxy.  As is apparent
from Figure~\ref{figcap5}, the inclination of the galaxy is small.  It
is well-known that an accurate measurement of {\it i} in face-on
galaxies is difficult \citep{walter01}.  Initial ROTCUR fits to all
parameters estimated {\it i} between $\sim$10\degree\ and
$\sim$40\degree, depending on the initial coordinates of the dynamical
center position.  We constrained the angle by forcing the fitted
rotation velocities (proportional to sin({\it i)}) to reach maxima
commensurate with the observed velocity field.  This narrowed the
allowable range of inclinations to 28\degree--32\degree; we fixed {\it
i} at 30\degree\ for the remainder of our analysis.

We then began fitting individual parameters of the rotation curve
using annuli of different widths (5\arcsec, 10\arcsec, 20\arcsec).
While the former two analyses do not provide independent points, they
do allow us to gauge the observed small-scale changes in parameters
that affect our analysis.  We compare values for a given parameter
using the fitted models at each resolution, and select a ``best-fit''
value that is consistent with all three fits.  In most cases, the
differences from one model to the next were negligible.  V$_{\rm sys}$
(3074\,$\pm$\,1 \kms) and the dynamical center position (11:38:45.3,
20:08:26.5) were fit individually using this approach, and then held
constant at all radii in subsequent model fits.  The P.A., which is
allowed to vary as a function of radius, was then fit with all other
parameters held constant.  Once values were obtained for each
parameter, a final rotation curve was derived with all of the above
variables held constant and rotation velocity (only) allowed to vary.

\begin{deluxetable*}{ccc}
\tabletypesize{\scriptsize}
\tablecaption{\HI\ vs. Optical Sizes in ``Giant Disk'' Dwarf Galaxies} 
\tablewidth{0pt}
\tablehead{
\colhead{Galaxy}         
&\colhead{$\mathcal{R}$(\HI/B)\tablenotemark{a}} 
&\colhead{Reference}}
\startdata
DDO\,154                        &21             &\citet{carignan89}, \citet{carignan98}\\
NGC\,2915                       &23\tablenotemark{b}               &\citet{meurer96}\\
UGC\,5288                       &27             &\citet{vanzee04}\\
NGC\,3741                       &39             &\citet{bremnes00}, \citet{begum05}\\        
ADBS 113845$+$2008	        &44 		&This work\\
\enddata
\tablenotetext{a}{Defined as the ratio of H~{\sc I}
size at the 10$^{19}$ cm$^{-2}$ level to the B-band exponential
scale length.}
\tablenotetext{b}{From \citet{meurer96}, quoted as the ratio of H~{\sc I} 
size at the 5\,$\times$\,10$^{19}$ cm$^{-2}$ level to the B-band
exponential scale length.}
\label{t2}
\end{deluxetable*}

\begin{figure*}
\plotone{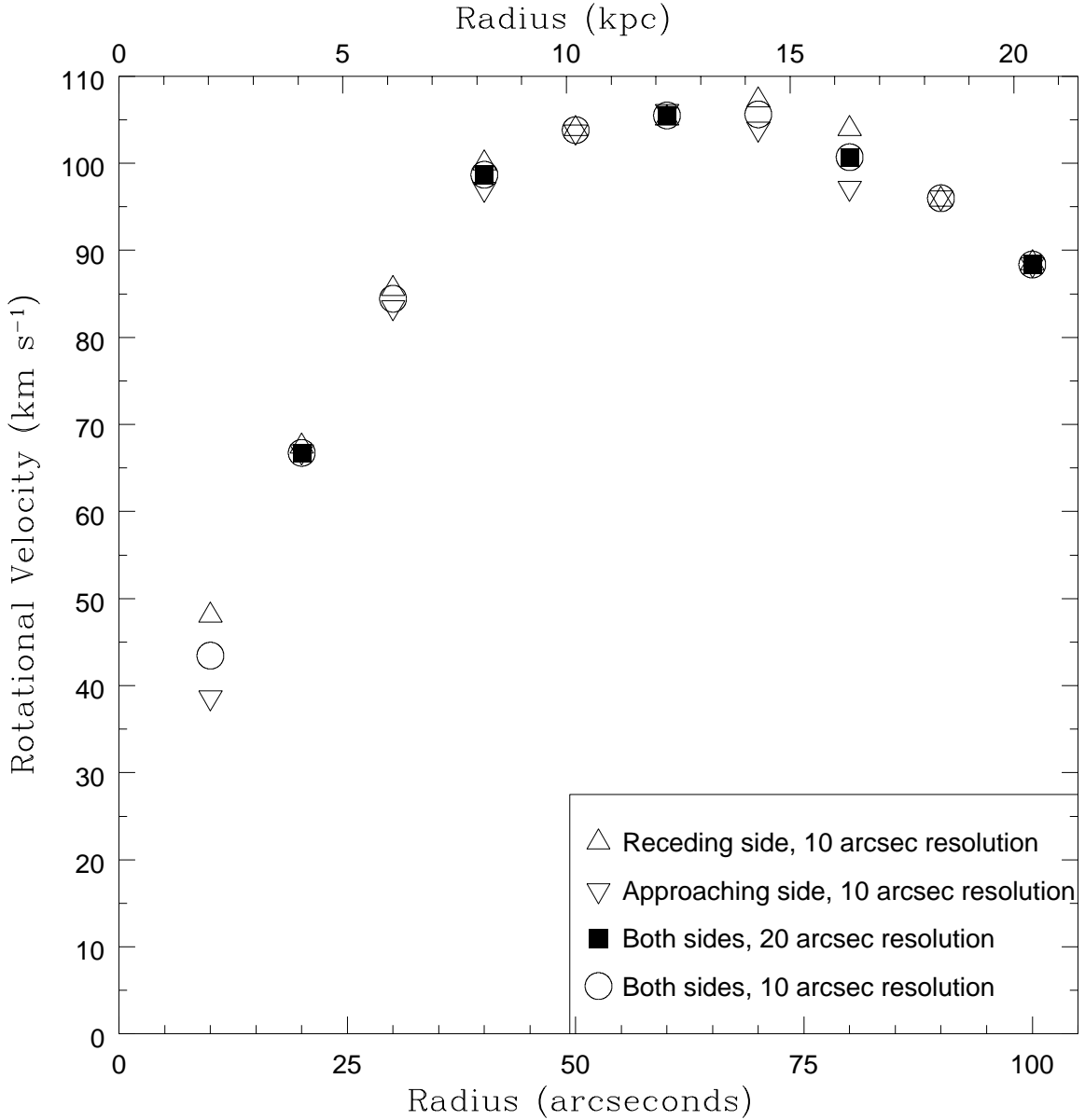}
\caption{Rotation curves of ADBS\,1138 extracted from the velocity
fields shown in Figure~\ref{figcap7}.  Fits to the approaching and
receding sides of the galaxy agree for most locations; the fits to
both sides average these discrepancies.  As expected from the
well-ordered rotation evident in Figure~\ref{figcap7}, the rotation
curve rises to $\sim$105 \kms\ in the inner 10 kpc and flattens out.
The turn-over of the outer rotation curve suggests that we have
sampled gas on physical sizes comparable to that of the galaxy's dark
matter halo.}
\label{figcap8}
\end{figure*}

The derived rotation curves are presented in Figure~\ref{figcap8}.
There we show two curves, derived using annuli of 10\arcsec\ and
20\arcsec, respectively.  We stress that only the lower resolution
points are completely independent (i.e., separated by a full beam
width).  The 10\arcsec\ curves agree with the 20\arcsec\ curves at all
radii and provide an indication of the rotation properties at radii
$<$4 kpc.  Note that we show curves derived for the receding and the
approaching sides individually, and the curve derived by averaging the
two sides together.  No significant differences were found in the
rotation curves of the two halves of the galaxy.  We discuss the
implications of the derived rotation curves in \S~\ref{S3.3}.

Inspecting the \HI\ moment zero images in Figure~\ref{figcap5}, it is
tempting to interpret the ring-like distribution of neutral gas as an
expanding shell.  To explore this possibility, we use the KARMA
visualization package to examine position-velocity and radius-velocity
diagrams of the \HI\ in \adbs.  Examining the kinematic information in
the radius-velocity plane has the advantage of integrating radially
across a putative shell structure in a datacube.  If a perfectly
symmetric, expanding shell were present in a given datacube, one would
see half of an ellipse in radius-velocity space.  If this radial
integration does not include a coherent, expanding structure, then one
will have \HI\ emission at a continuous range of velocities as a
function of radius (i.e., the ``ellipse'' will be closed).

\begin{figure*}
\plotone{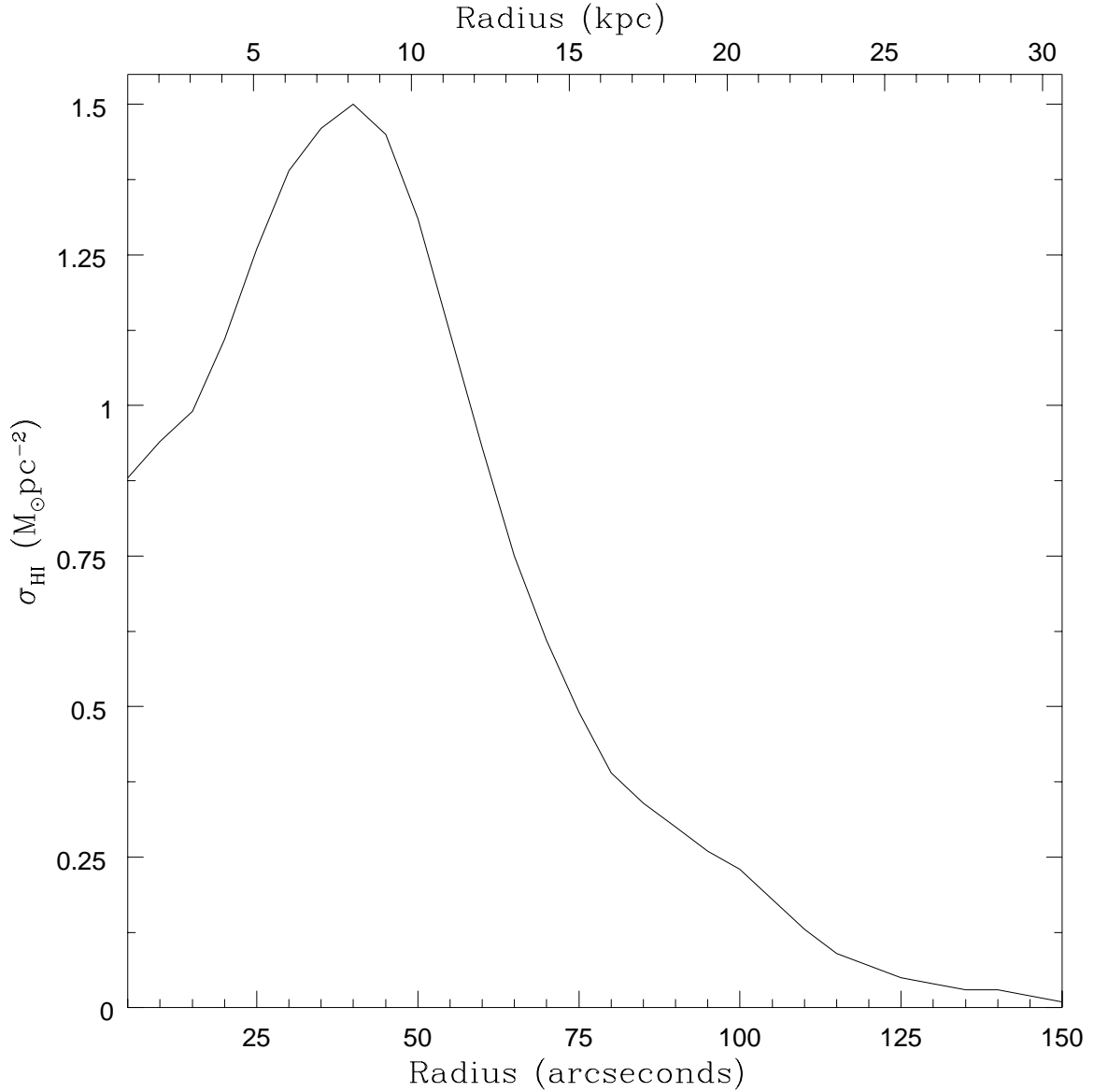}
\caption{Radially averaged surface brightness profile of ADBS\,1138,
created by summing \HI\ emission in concentric rings emanating from
the dynamical center found in our rotation curve analysis.  This
profile provides clear evidence for the central depression of \HI\ and
the ring structure that dominates the galaxy's \HI\ morphology.} 
\label{figcap6}
\end{figure*}

\begin{figure*}
\plotone{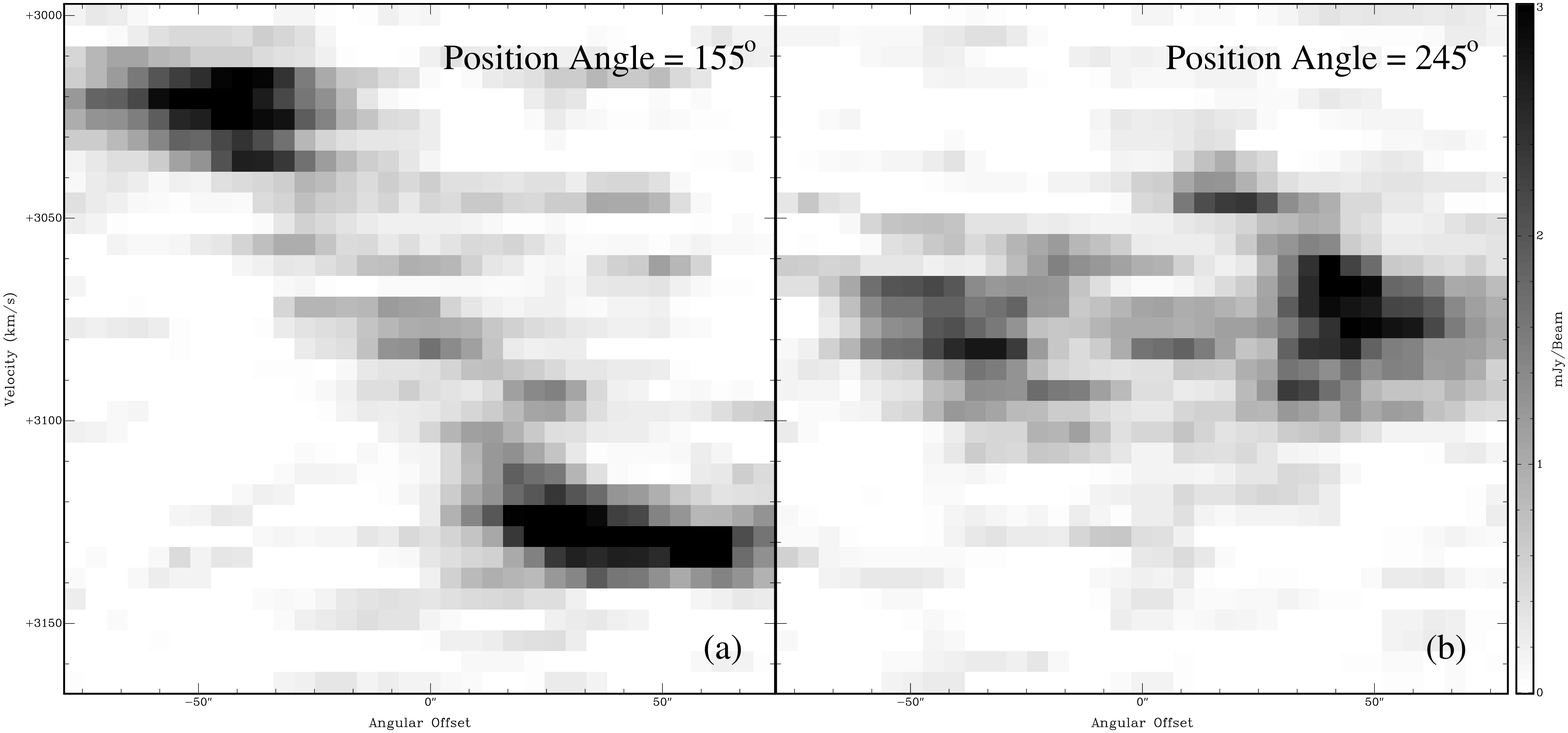}
\vspace{1 cm}
\caption{\HI\ major (a) and minor (b) axis position-velocity cuts
through the dynamical center of the \adbs\ system.  The central
depression and surrounding ring structure are again prominent
features.  Integrating emission radially from the dynamical center
position does not show the unambiguous signs of expansion.}
\label{figcap9}
\end{figure*}

From Figures~\ref{figcap5} and \ref{figcap6} it is obvious that a broken,
roughly spherical structure exists in the ISM of \adbs\ at a distance
of $\sim$7.5 kpc from the dynamical center.  However, our kinematic
analysis of this structure does not show the unambiguous signs of
expansion.  A major-axis (position angle $=$
155\degree, measured East of North) position-velocity diagram (see
Figure~\ref{figcap9}) does show the \HI\ column density enhancement at
this radius, with mostly coherent \HI\ gas at intermediate radii.
Similarly, a minor-axis (position angle $=$ 245\degree)
position-velocity cut (see also Figure~\ref{figcap9}) shows higher
columns at the same radius, with gas at roughly the same velocity from
one side of the galaxy to the other (as expected, based on the
well-behaved velocity field in the inner regions of the galaxy; see
Figure~\ref{figcap7}).  Radius-velocity diagrams do not show the
characteristic elliptical structure that would be expected if the \HI\
ring was in fact expanding along the line of sight.  We note that this
galaxy is observed at a small inclination ({\it i} was fixed at
30\degree\ in the rotation curve analysis above); for nearly face-on
galaxies there might be structures that are expanding in the disk
(perpendicular to the line of sight) but whose mass outflow
perpendicular to the disk is small (hence undetectable).

%-----------------------------------------------------------------------------%
\subsection{Dark Matter in ADBS 113845+2008}
\label{S3.3}
%-----------------------------------------------------------------------------%

The rotation curves in Figure~\ref{figcap8} show ordered rotation in
the central regions.  This behavior continues out to a radius of
$\sim$50\arcsec\ ($\sim$10 kpc), at which point there is a pronounced
change to differential rotation at larger radii.  Using these curves
it is straightforward to estimate the dynamical mass of the system.
Assuming the rotational velocities shown are for neutral gas on
circular orbits, we find that the dynamical mass included within
50\arcsec\ ($\sim$10 kpc; V$_{rot}$ $\simeq$ 100 \kms) is
$\sim$2.3\,$\times$\,10$^{10}$ \msun.  At the last measured point of
the rotation curve (r $=$ 100\arcsec\ $\simeq$20 kpc, V$_{rot}$
$\simeq$ 90 \kms) the dynamical mass increases to
$\sim$3.8\,$\times$\,10$^{10}$ \msun.  We estimate the uncertainties
on these dynamical mass values to be $\sim$20\%.

As derived in \S~\ref{S3.1}, the total \HI\ gas mass is
(1.08\,$\pm$\,0.22)\,$\times$\,10$^{9}$ \msun.  Accounting for
primordial Helium increases the neutral gas mass to
(1.46\,$\pm$\,0.30)\,$\times$\,10$^{9}$ \msun.  Due to the metal-poor
nature of the ISM in \adbs, and the known problems of measuring
molecular gas masses in such conditions, we make no explicit
correction for such material and consider the \HI\ $+$ He to
constitute all of the cool gas in \adbs.

A stellar mass estimate is needed to calculate the total luminous mass
and to infer the ratio of dark to luminous matter.  Using our {\it
WIYN} 0.9-m photometry (M$_{\rm B}$ = $-$16.71, M$_{\rm V}$ =
$-$17.30, M$_{\rm R}$ = $-$17.67, after foreground extinction
corrections) and the models presented in \citet{bell01}, we can infer
stellar mass-to-light ratios for a range of galaxy and stellar
population characteristics.  Unfortunately we do not have infrared
imaging of \adbs\ that would extend the color baseline and reduce
scatter due to differential extinction and recent star formation (see
the discussion in {Bell \& de~Jong 2001}\nocite{bell01} and
application in {Lee \etal\ 2006}\nocite{lee06}).  As such, we stress
that our stellar mass is only roughly constrained at this time.

We estimate the stellar mass of ADBS\,1138 as follows.  First, the
extinction corrected colors and luminosities are calculated: (B$-$R)
$=$ 0.96; (B$-$V) $=$ 0.59; L$_{\rm V}$ $=$ 7.02\,$\times$\,10$^8$
L$_{\odot}$; L$_{\rm R}$ $=$ 6.03\,$\times$\,10$^8$ L$_{\odot}$.
Next, we calculate the stellar mass-to-light ratio and the stellar
mass for multiple metal-poor model sets in \citet{bell01}.  Averaging
these results for each of the (B$-$R) and (B$-$V) colors, and using
the dispersion among them as an indicator of uncertainty, we
approximate the stellar mass of ADBS\,1138 to be
(9.9\pom5.0)\,$\times$\,10$^8$ \msun\ [using the (B$-$R) colors] and
(1.4\pom0.7)\,$\times$\,10$^9$ \msun\ [using the (B$-$V) colors].
Since the (B$-$R) color baseline is larger (see above), we adopt a
stellar mass of (1\pom0.5)\,$\times$\,10$^9$ \msun\ for \adbs.
Repeating this procedure with colors derived from {\it SDSS}
photometry, we find excellent agreement with this value (well within
the errorbars).

Combining our total gas mass measurement (see above) and this (coarse)
stellar mass estimate, we find that the dynamical mass exceeds the
luminous mass by a factor of 9 (at a radius of 10 kpc) to 15 (at the
last measured point of the rotation curve).  Like many dwarf galaxies,
ADBS\,1138 is a dark matter dominated system.  We apparently have
sampled the turn-over of the outer rotation curve (see
Figure~\ref{figcap8}); this suggests that we have measured the bulk of
the galaxy's dark matter halo.

%-----------------------------------------------------------------------------%
\section{The Recent Evolution of ADBS 113845+2008}
\label{S4}
%-----------------------------------------------------------------------------%

\adbs\ presents intriguing problems: 1) How can star formation proceed
at the current time in regions far below the ``canonical'' star
formation threshold column density of $\sim$10$^{21}$ cm$^{-2}$?  2)
What gives rise to a giant ordered disk that only forms stars in the
very central region?  In this section we discuss two potential
evolutionary scenarios for this system.  In the first we consider
``inside-out'' processes: here we assume that the structures we
observe in the neutral gas disk are a result of star formation
processes.  In the second we consider ``outside-in'' scenarios, where
we assume that the large \HI\ disk has been stable for a significant
fraction of a Hubble time, and that the recent star formation we
observe is a result of small-scale stochastic effects that caused the
neutral gas to attain critical density only in the very innermost
regions.

%-----------------------------------------------------------------------------%
\subsection{``The Inside-Out Evolutionary Scenario''}
\label{S4.1}
%-----------------------------------------------------------------------------%

We first consider the ``inside-out'' scenario wherein the ring-like
\HI\ distribution is the result of ``feedback'' from recent star
formation.  If a recent star formation episode was sufficiently
energetic, one might expect the mechanical luminosity from stellar
winds and SNe to be driving \HI\ away from the stellar component.
Numerous nearby dwarf galaxies have interstellar media displaying such
kinematic structures (e.g., {Puche \etal\ 1992}\nocite{puche92}, {Kim
\etal\ 1999}\nocite{kim99}, {Walter \& Brinks 1999}\nocite{walter99},
{Ott \etal\ 2001}\nocite{ott01}, and references therein).

The depressed column densities around the optical component (compared
to ``canonical'' expectations based on the ongoing star formation; see
references above), coupled with the apparent ``post-starburst'' nature
of the galaxy, make this scenario appealing.  The compact stellar
component implies that all mechanical energy input from previous
bursts would be highly concentrated spatially; \adbs\ appears to
possess the necessary conditions for the creation of a large \HI\
shell in a low mass galaxy.  Here, we can exploit the kinematic
information available in our spectral line data to discern if the
neutral gas is in fact being driven outward from the galaxy. 

It would be instructive to consider the energetics necessary to create
such a coherent structure in the ISM.  Ideally we would have the
signature of expansion of the ring; using this velocity and the
observed size of the structure one can obtain a first-order estimate
of the requisite energy and age from basic first principles.  However,
in the case of \adbs, we do not detect expansion and thus have no such
estimate of the velocity; this relegates us to a first-order estimate
of the ring energetics.  We begin by assuming that the extended \HI\
envelope of \adbs\ existed with more or less its current physical size
at the beginning of a putative (presumably violent) starburst episode
that was concentrated in the central region (coincident with the
optical component).  The low column densities throughout the extended
disk are not unlike those found in the outer disks of nearby dwarf
galaxies with extended \HI\ envelopes \citep{hunter98a}.  We then
assume that the mass of gas in the ring was distributed uniformly
throughout the central region of the galaxy, presumably providing
plentiful neutral gas that gave rise to the starburst event inferred
from the spectrum shown in Figure~\ref{figcap1}.

As shown in \S~\ref{S3.1}, the mass of gas in the \HI\ ring is
(5.2$\pm$0.6)\,$\times$\,10$^{8}$ \msun.  Assuming a symmetric
potential interior to and inclusive of the ring, this mass has a
potential energy of $\sim$10$^{54}$ erg at the present time.  For this
gas to be moved radially outward a distance of $\sim$7.5 kpc as a
result of spatially and temporally concentrated star formation, the
kinetic energy inputs of some thousands of SNe would be required.
Given a characteristic starburst duration of $\sim$100 Myr
\citep{cannon03,weisz08,mcquinn09}, a constant-velocity expansion rate
of $\sim$70 \kms\ is inferred; a longer-duration burst would decrease
the constant expansion velocity.  

It is interesting to note that expanding shells in other galaxies have
been found that require energies of roughly the same order as those
inferred here \citep[e.g.,][]{ott01}.  However, this interpretation is
seriously weakened by the lack of observed expansion in the ring
itself.  Of course it is possible that the kinetic energy of the
once-expanding ring has been expended, that pressure equilibrium with
the surrounding ISM has been reached, and that the expansion has
stopped.  Without a definitive measurement of expansion we must estimate
the total mechanical energy input from previous star formation and
subsequent stellar evolution.  Coupled with the enormous size of the
ring compared to the stellar component, we conclude that the ring
structure in ADBS\,1138 was most likely not created by concentrated
star formation alone.

We again draw attention to the size of the ring structure; the radius
of 7.5 kpc greatly exceeds the physical sizes of some of the largest
\HI\ structures that have been attributed to stellar evolution
processes (e.g., NGC\,6822 - {de~Blok \& Walter
2000}\nocite{deblok00}; Holmberg\,I - {Ott \etal\
2001}\nocite{ott01}).  Examples of ``ring-like'' galaxies have been
identified in the literature, either on the basis of a
visually-obvious depression and surrounding ring of higher column
densities (e.g., {Simpson \etal\ 2005}\nocite{simpson05}) or on the
basis of radially-integrated \HI\ surface density profiles (e.g.,
{Walter \etal\ 2007}\nocite{walter07}).  Centrally-concentrated star
formation is argued to produce the observed structures in some of
these galaxies, even in cases where no expansion is detected.  Thus,
what makes \adbs\ remarkable is not the rarity of ring structures in
dwarf galaxies, but rather the comparative sizes of the optical and
gaseous components.  We stress that the optical component of \adbs\
(measured by the B-band exponential scale length) is more than 40
times smaller than the extended \HI\ halo (measured as the average
radius at which the column density falls to the 10$^{19}$ cm$^{-2}$
level).  Ongoing star formation is detected only in the compact
optical component.

%-----------------------------------------------------------------------------%
\subsection{``Outside-In'' Evolutionary Scenario}
\label{S4.2}
%-----------------------------------------------------------------------------%

The shortcomings noted above lead us to consider an alternative,
``outside-in'' scenario for the recent evolution of \adbs.  We
postulate that the extended, relatively tenuous (i.e., low column
density) \HI\ disk has remained stable for essentially a Hubble time,
with star formation occurring only in very localized regions.  In most
respects the \HI\ disk in \adbs\ is like the \HI\ disks typically
associated with low-mass galaxies: coherent and ordered rotation, with
sub-critical column densities throughout most of the disk.  Few
locations within this extended neutral gas halo have attained the
requisite gas densities to initiate star formation; at present, the
only such location is the very innermost region of the disk.  Given
the very compact stellar component in the optical bands, it follows
that star formation has likely been concentrated in this very inner
region over the lifetime of the galaxy.

However, at this physical resolution, we do not detect high column
density \HI\ gas at the location of the optical component (though
molecular gas may be present; see discussion below).  We thus
postulate that localized star formation has been efficient at
depleting the neutral gas in this central region.  Whatever \HI\ gas
that gave rise to the present (and past) star formation at this
location is either very spatially concentrated (and thus undetected
due to beam smearing effects at our resolution) or has been consumed
in the star formation process.  In contrast, the vast majority of the
disk is made up of tenuous, sub-critical gas undergoing ordered
rotation that is not associated with ongoing star formation.  We thus
further postulate that the global star formation has been exceedingly
inefficient.

While the efficiency with which stars are formed in the ISM has been
long-studied \citep[e.g.,][]{thronson86}, it is difficult to state
with certainty the ``average'' efficiency of star formation in dwarf
galaxies.  \citet{vanzee97} find differences in the implied star
formation efficiencies between a sample of low surface brightness
dwarfs and ``normal'' gas-rich dwarfs, but no apparent distinction
between the two samples based on the strength of current star
formation.  Thus, star formation in dwarfs appears to be a stochastic
process: where dense neutral gas is located, stars will form.  The
physical mechanism(s) that initiate(s) this process are widely
debated.  Two of the most common drivers speculated to be operating in
dwarfs are interactions and turbulence; we briefly discuss both
mechanisms below.

Galaxy-scale interactions are widely postulated to compress gas in
galaxies, thereby triggering starbursts
\citep[e.g.,][]{barnes92,mihos96,sanders96}.  In massive disk galaxies
in group environments, this process is frequent; this is a
well-studied driver of galaxy evolution.  However, in low-mass
galaxies, the role of interactions is more difficult to quantify.  We
know that interactions of dwarfs with massive systems leads to
disruption and accretion of the dwarf by the more massive system
\citep[e.g.,][]{mihos94}.  However, the frequency and results of
mergers or interactions between two low-mass systems are not
constrained observationally.  Assuming that such scenarios occur
(requiring either close pairs of galaxies, or neutral gas clouds of
roughly dwarf galaxy mass interacting with dwarfs), it is logical to
assume similar impacts in the low-mass galaxies as those seen in the
more massive systems.

This scenario is attractive in that it provides a simple explanation
for the observed properties of the \adbs\ system.  In such a model, at
some time in the past, a low-mass object underwent a strong
interaction with the neutral gas disk of this system.  The interaction
drives gas into the inner disk, at which point critical gas density is
achieved and stars begin to form in earnest.

At the present time, \adbs\ appears to be an isolated galaxy.
Searching our ``low-resolution'' datacube we find no (gas-rich)
companion or interacting galaxy at our current sensitivity level.  We
note that the {\it VLA} primary beam at 21\,cm ($\sim$32\arcmin)
corresponds to a circular field size of radius $\sim$195 kpc at our
adopted distance.  Any relatively massive (M$_{\rm HI}$ $\sim$10$^7$ \msun)
companion within this projected separation would have been detected in
our data.  If such an interaction scenario is correct, then the lack
of a (gas-rich) companion might be explained by either the process
occurring long ago (allowing the \HI\ disk time to recover stability)
or by the companion being completely accreted in the central regions
of the galaxy.

Alternatively, turbulent motions in the ISM of \adbs\ may have
initiated and sustained star formation in the inner disk.  It is
widely suggested that the ISM of dwarf galaxies are fractal in nature,
with turbulence playing an important role in regulating the star
formation process
\citep[e.g.,][]{vogelaar94,elmegreen98,rhode99,westpfahl99,maclow04,joung06}.
However, we also note that direct comparisons of simple turbulent
models with observations sometimes are more successful in massive
spirals than in irregulars \citep{hunter98b}.

If the inner disk of \adbs\ contains substantial turbulent motion,
then this conceivably could lead to ongoing star formation.
``Feedback'' from this star formation will unleash mechanical energy
into the ISM, potentially contributing to the turbulent nature of the
ISM and thereby sustaining the star formation within the disk.  Given
the difficulties of completely removing the ISM from dwarf galaxies
that are undergoing star formation events \citep[e.g.,][]{maclow99},
such a scenario appears to be a viable (though overly simplistic)
model of the evolution of \adbs.

We note that there is likely a molecular component of the ISM to which
we are not sensitive with these data.  Moving inward through the disk,
the ISM may simply become predominantly molecular inside of $\sim$ 7.5
kpc; examples of this transition are numerous
\citep[e.g.,][]{walter08}.  This may provide a logical explanation for
the ring structure and for the depressed column densities in the
central region.  This putative molecular gas must still be relatively
tenuous, since we only observe ongoing star formation in one
relatively small region within the ring.  While there may be
widespread molecular gas inside the ring, only the innermost $\sim$1
kpc appears to have had sufficient densities to support star formation
within the past $\sim$20 Myr.  Based on the nebular abundance
($\sim$30\% Z$_{\odot}$), and the known problems of studying molecular
gas at low metallicities \citep[e.g.,][]{taylor98}, studying the
morphology and kinematics of this molecular gas will be difficult (see
also the study by {Leroy \etal\ 2005}\nocite{leroy05} and references
therein).  Considering the low dust contents typically associated with
metal-poor dwarfs \citep[e.g.,][]{draine07,engelbracht08}, it is
unlikely that \adbs\ hosts undetected or embedded star formation
(although examples have been identified; see {Cannon \etal\
2006}\nocite{cannon06}).

We wish to stress that we do not advocate for the youth of \adbs; we
have no evidence that this is a ``primordial'' galaxy.  On the
contrary, we see evidence for an older stellar population in our
spectrum.  Unfortunately, the large distance to \adbs\ ($\sim$42 Mpc)
precludes resolved stellar population work that would provide
characteristics of this older stellar population.  Rather than being a
``young'' galaxy, we postulate that \adbs\ is an apparently rare
system in the local universe; the large \HI\ disk is very stable and
shows no signs of recent tidal interaction.  The star formation in the
very innermost region of the disk (where the red stellar population is
located) was stronger in the past.

Taken as a whole, the lines of evidence provided by \HI\ spectral line
and optical imaging paints a very intriguing picture.  \adbs\ is a
relatively massive system compared to other dwarfs
\citep[e.g.,][]{mateo98}.  The tenuous \HI\ appears to be very stable
and to be happily evolving in isolation at the present time.  In the
very innermost region, the requisite conditions for star formation
were met; we speculate that a low-mass merger event or turbulent
processes may have been important in initiating and/or regulating this
star formation.  We favor an ``outside-in'' scenario to explain the
observed features of \adbs; while this model is simplistic and has
shortcomings, it appears to be in better agreement with the data than
the ``inside-out'' models discussed in \S~\ref{S4.1}

%-----------------------------------------------------------------------------%
\section{Conclusions}
\label{S5}
%-----------------------------------------------------------------------------%

We have presented new optical imaging and spectroscopy and \HI\
spectral line imaging of the intriguing BCD galaxy \adbs.  This system
displays extreme characteristics that differentiate it from the
``typical'' dwarf galaxy in the local universe.  First, from optical
imaging, this system harbors ongoing star formation in an extremely
compact stellar distribution.  The B-band exponential scale length is
only 0.57 kpc. While more compact systems have been found, both among
the BCD and the ``ultra-compact dwarf'' populations, the latter are
typically ``threshed'' systems that are the remnant nucleii of
disrupted dwarf galaxies \citep{drinkwater03}.  \adbs\ is one of the
most optically compact, isolated, star-forming dwarf galaxies known to
date.

The compact optical component is in stark contrast to the extremely
extended neutral gas disk.  The \HI\ disk is $\sim$44 times larger
than the B-band exponential scale length; this makes \adbs\ the most
\HI-extended galaxy known. If a low surface brightness stellar
component exists, it must be very faint indeed; we find very little
variation of optical scale length with {\it SDSS} observing band.
Sensitive mid-infrared imaging would be insightful to probe the total
stellar extent of the system.

The \HI\ morphology is dominated by a broken ring $\sim$7.5 kpc in
radius, centered on the optical component.  Nowhere in the galaxy do
\HI\ column densities rise above the ``canonical'' star formation
surface density threshold of 10$^{21}$ cm$^{-2}$ (though beam smearing
effects may be important).  Rather, the optical component occupies
only the very innermost portion of an otherwise stable, extended \HI\
disk.  The resolution of our data (20\arcsec\ = 4.08 kpc) precludes a
detailed, spatially resolved Toomre Q analysis
\citep[e.g.,][]{martin01}.  However, recent studies of nearby galaxies
have shown close agreement between \HI\ column density and Toomre Q
parameter as predictors of locations of active star formation
\citep[e.g.,][]{deblok06}; higher spatial resolution observations would
allow more detailed analysis of the stability of the large \HI\ disk
in \adbs.

We investigate the kinematics of this ring structure in both
position-velocity and radius-velocity space.  We find no clear
signatures of large-scale expansion of this feature.  While very
coarse estimates of the energy required to create the ring are
available from simple potential arguments, we are unable to
definitively conclude that coherent star formation and subsequent
stellar evolution is responsible for the creation of this ring
feature. Based on optical spectra that show ``post-starburst''
characteristics, this scenario is appealing: a compact stellar
distribution would imply spatially concentrated energy input by
massive star evolution.  However, given the physical size and the
kinematically static nature of the ring, we seek an alternative
explanation for the ring's origin.

We consider evolution in the ``outside-in'' sense, where the \HI\ disk
has been stable for essentially a Hubble time, and where localized,
stochastic effects have initiated the growth of the stellar
population.  We speculate on two possible mechanisms of this ongoing
star formation: minor-minor merger/interaction or interstellar
turbulence.  While the distance of \adbs\ precludes resolved stellar
population work that would more accurately constrain the age of the
stellar population, we infer that this is not a young galaxy.  Rather,
we have discovered an apparently rare object that appears content to
evolve in quiescent isolation; inefficient global star formation has
produced a system with a very concentrated stellar population but with
an enormously extended \HI\ disk.

Within the inner disk, the neutral gas is undergoing the ordered
rotation that is typical of many dwarf galaxies.  We sample the
rotation curve past the point where it turns over, indicating that we
have determined the size of the dark matter halo of this system.
Simple arguments yield a total dynamical mass of
(2--4)\,$\times$\,10$^{10}$ \msun\ (depending on the selected distance
from the dynamical center).  This outweighs the luminous matter (gas
$+$ stars, without an explicit estimate of the molecular gas mass) by
at least a factor of 15.

The curious properties of \adbs\ raise important questions.  What
fraction of emission-line dwarfs display such compact optical bodies?
Are large \HI\ halos common amongst such systems, or is \adbs\ unique?
How are large, quiescent \HI\ disks formed and maintained?  Are the
low column densities found in \adbs\ typical of those in compact
dwarfs?  Is the neutral-to-molecular transition more efficient in
certain regions of these systems than in more extended dwarfs?  Are
ring structures in the ISM indicative of this transition, do they
arise from the combined effects of stellar evolution, or do they
signify another (as yet unidentified) process that shapes the ISM?  A
detailed study of the gaseous components of a more complete sample of
optically compact systems would be exceptionally fruitful in
addressing these important questions.

%-----------------------------------------------------------------------------%
\acknowledgements

JMC thanks the National Radio Astronomy Observatory and Macalester
College for partial support of this work.  The authors thank Henry
Lee, Gustaaf van Moorsel, Liese van Zee and Fabian Walter for helpful
discussions.  This research has made use of the NASA/IPAC
Extragalactic Database (NED) which is operated by the Jet Propulsion
Laboratory, California Institute of Technology, under contract with
the National Aeronautics and Space Administration, and NASA's
Astrophysics Data System.

%-----------------------------------------------------------------------------%
%\clearpage
%\bibliographystyle{apj}                                                 
%\bibliography{references,unread}   

%-----------------------------------------------------------------------------%
\end{document}